\documentclass[10pt]{IEEEtran}  %,draftcls,peerreview

\usepackage{xr} % cross references to external docs. http://www.tex.ac.uk/cgi-bin/texfaq2html?label=extref
\usepackage{amsmath,amsthm, amssymb, epsfig, verbatim, color, amsfonts} % spconf
\usepackage{graphicx}
\ifx\theorem\undefined
\theoremstyle{plain}
\newtheorem{theorem}{Theorem}
\newtheorem{lemma}{Lemma}%[theorem] % enable the remark in order to number the lemmas as subnumbering of theorems
\newtheorem{corollary_in_theorem}{Corollary}[theorem]
\theoremstyle{definition}
\fi

\def\vr{\mathbf}
\def\E{\mathbb{E}}

\def\Ind{\mathrm{Ind}} % another option: uspackage{bbm}, \mathbbm{1}

\def\half{\tfrac{1}{2}}
\def\Var{\mathrm{Var}}
\def\endofproof{\hspace{\stretch{1}}$\Box$}
\def\defeq{\triangleq} % \equiv

\def\unif{\mathbb{U}}

\newcommand{\tsubs}[1]{{\scriptscriptstyle \mathrm{#1}}}
\def\ntoinfty{\arrowexpl{n \to \infty}}

\newcommand{\argmin}[1] {\underset{#1}{\textstyle \mathrm{argmin}}\hspace{0.5ex}}
\newcommand{\arrowexpl}[1] {\underset{#1}{\textstyle \longrightarrow}}

\ifx\note\undefined
\newenvironment{note}{\comment}{\endcomment} % uncomment this line in order to remove all "notes" paragraphs
\fi
\newenvironment{inputpath}[1]
{ \let\origtexinput\input \renewcommand{\input}[1]{\origtexinput{#1/##1}}
\let\origtexincludegraphics\includegraphics \renewcommand{\includegraphics}[2][]{\origtexincludegraphics[##1]{#1/##2}} }
{ \let\input\origtexinput \let\includegraphics\origtexincludegraphics}

\allowdisplaybreaks[1]

\def\bbest{\mathbf{B}}
\def\Nfpf{N^\tsubs{(p)}}
\def\Lfpf{L^\tsubs{(p)}}
\def\Ncl{N^\tsubs{(cl)}}
\def\Lcl{L^\tsubs{(cl)}}

\newcommand{\onlyfull}[1]{#1}
\newcommand{\onlyconf}[1]{}
\newcommand{\excluded}[1]{}

\title{A Universal Probability Assignment for Prediction of Individual Sequences}

\author{Yuval Lomnitz, Meir Feder \\
Tel Aviv University, Dept. of EE-Systems  \\
Email: yuval.lomnitz@gmail.com ,meir@eng.tau.ac.il}

\begin{document}
\maketitle

\begin{abstract}
Is it a good idea to use the frequency of events in the past, as a guide to their frequency in the future (as we all do anyway)? In this paper the question is attacked from the perspective of universal prediction of individual sequences. It is shown that there is a universal sequential probability assignment, such that for a large class loss functions (optimization goals), the predictor minimizing the expected loss under this probability, is a good universal predictor. The proposed probability assignment is based on randomly dithering the empirical frequencies of states in the past, and it is easy to show that randomization is essential. This yields a very simple universal prediction scheme which is similar to Follow-the-Perturbed-Leader (FPL) and works for a large class of loss functions, as well as a partial justification for using probabilistic assumptions.
\end{abstract}

%\begin{IEEEkeywords}
%\end{IEEEkeywords}

\section{Introduction}
In this paper the problem of universal sequential prediction of an individual unknown sequence is considered \cite{Nicolo}\cite{FederMerhav93}\cite{FederMerhav98}, and a prediction approach based on universal probability assignment is proposed. Given a space of strategies $\mathcal{B}$, a space of nature states $\mathcal{X}$ and a loss function $l(b,x), b \in \mathcal{B}, x \in \mathcal{X}$, the purpose is to assign the next strategy $\hat b_t$ given the knowledge of the past states $\vr x_1^{t-1}$, such that the overall loss $\sum_{t=1}^n l(\hat b_t, x_t)$ would be asymptotically close to the loss obtained by the best fixed strategy known a-posteriori after viewing the entire sequence $\vr x_1^n$, i.e. $\min_{b \in \mathcal{B}} \sum_{t=1}^n l(b, x_t)$.

In the particular case of sequential probability assignment under the $\log$ loss function $l(b,x) = \log \frac{1}{b(x)}$ where $\mathcal{B}$ is the space of probability assignments on the finite alphabet $\mathcal{X}$, or equivalently in universal sequential compression, it is shown \cite{FederMerhav98}\cite[\S13]{CoverThomas_InfoTheoryBook}\cite[\S9]{Nicolo} that it is possible to assign probabilities $\hat p_t(x_t)$ for the next state in an arbitrary sequence of states $x_t \in \mathcal{X}, t=1,2,\ldots, n$, given the past states, such that for any possible sequence, the overall probability $\hat p(\vr x) = \prod_{t=1}^n \hat p_t(x_t)$ would not be too far, in a multiplicative or logarithmic sense, from the best i.i.d. probability assigned to the sequence a-posteriori $\max_{p(\cdot)} \prod_{t=1}^n p(x_t)$. The result extends to probability assigned by Markov machines or finite state machines \cite{Feder_Gambling91}. This problem is related to universal compression because the overall compression length corresponds to $\log \left( \frac{1}{p(\vr x)} \right)$. A remarkable feature of these universal probability assignments is that, although nothing is assumed about the sequence, to construct a universal encoder it is enough to encode as if $\hat p_t(\cdot)$ was the \emph{true} probability of the next state.

These universal probability assignments, such as the Laplace \cite[\S13.2]{CoverThomas_InfoTheoryBook} or Krichevsky-Trofimov (KT) \cite{KrichevskyTrofimov81} estimators, have an intuitively appealing structure which induces a small bias over the empirical distribution seen so far. For example, Laplace's estimate for the probability distribution of of $x_t$ is
\begin{equation}\label{eq:54}
\hat p_t(x) = \frac{N_{t-1}(x) + 1}{(t-1) + |\mathcal{X}|}
,
\end{equation}
where $N_t(x)$ denotes the number of times the state $x$ appears in $\vr x_1^t$. While these estimators get closer with time to the measured empirical distribution, they do not ``trust'' it completely, and, for example, never assign a probability value $0$ to states that had not appeared before. Furthermore, in the probabilistic prediction setting the same distributions were shown to perform well not only for the $\log$ loss: the predictor which minimizes the expected loss under these distributions $\hat b_t = \argmin{b} \underset{X \sim \hat p_t(\cdot)}{\E} l(b,X)$ operates well for a wider class of loss functions \cite[\S III.A.2]{FederMerhav98}.

This naturally leads to the following question: is it possible to forecast an individual sequence by first generating a probability assignment based on the past, and then minimizing the expected loss under this assignment (i.e. in a way, acting as if future events truly happen with this probability)? Consider prediction schemes of the following form:
\begin{enumerate}
\item Generate a probability assignment $P^{(u)}_t(x)$ based on the past of the sequence $\vr x_1^{t-1}$, in a way which does not depend on the loss function.
\item To predict $b_t$ under the loss function $l(b,x)$, choose the strategy that minimizes the expected loss under $P^{(u)}_t$, i.e.:
\begin{equation}\label{eq:70}
\hat b_t = \argmin{b} \underset{X \sim P^{(u)}_t(\cdot)} \E \left[ l(b,X) \right]
\end{equation}
\end{enumerate}
If there exists a single scheme for generating $P^{(u)}_t(\cdot)$ that does not depend on the loss function $l(b,x)$, but for which $\hat b_t$ yields a good (Hannan-consistent \cite{Nicolo}) predictor for a certain class of loss functions, then we call $P^{(u)}_t(\cdot)$ a \textit{universal sequential probability assignment} with regards to that class. Notice that this term has been used in the past with respect to the $\log$-loss, so the definition above can be considered a natural extension.

It is easy to show that, if the class of loss functions includes even simple loss functions such as the 0-1 loss (the number of errors), then no deterministic assignment can be universal, and therefore the Laplace or KT assignments are inadequate. However, it is shown in this paper that the random assignment obtained by slightly perturbing the empirical frequencies is universal for a large class of loss functions, including the $\log$-loss and any bounded loss.

In addition to supplying a simple and general universal prediction scheme, this result also has interpretations contributing to our understanding of probability. For example, it supplies justification for treating the statistics of a process in the past as a guide to its statistics in the future, without having to assume the process is indeed stationary, or that it is driven by a ``probabilistic'' law. In other words, if our natural behavior is in some way similar to the prediction algorithm described here, then the claims on its convergence can be used to justify this behavior.

The next section completes the problem definition and discusses the boundaries of the solution, and relations to known results. Section~\ref{sec:main_results} gives the main
results\onlyconf{~(the proofs are omitted and can be found in the full paper \cite{YL_FPF})}, and Section~\ref{sec:philo_part} discusses the possible implications on understanding probabilistic behavior. \onlyfull{The proofs are given in Section~\ref{sec:proofs}.}

\section{Problem statement and discussion}\label{sec:problem_statement}
Building upon the definitions already presented in the introduction, in this section some complementary definitions are presented. We assume throughout this paper that $\mathcal{X}$ is finite (otherwise there is no meaning to measuring empirical frequencies). The set of possible strategies $\mathcal{B}$ is not restricted. The loss function $l(b,x)$ is constant over time.

Let us define the accumulated loss of a sequential predictor $\hat b_t(\vr x_1^{t-1})$ as:
\begin{equation}\label{eq:96}
\hat L_n = \sum_{t=1}^n l(\hat b_t, x_t)
,
\end{equation}
and the loss of the best fixed strategy as:
\begin{equation}\label{eq:104}
L^*_n = \min_b \sum_{t=1}^n l(b, x_t)
.
\end{equation}
The difference $\hat L_n - L^*_n$ which is defined as the regret, is a function of the predictor and the sequence. The worst case regret is:
\begin{equation}\label{eq:109}
\mathcal{R}_\tsubs{max} = \max_{\vr x_1^n} \left( \hat L_n - L^*_n \right)
,
\end{equation}
and the normalized regret is $\frac{\mathcal{R}_\tsubs{max}}{n}$. A forecasting strategy $\hat b_t$ is said to be \textit{Hannan-consistent}, if  $\limsup_{n \to \infty} \frac{\mathcal{R}_\tsubs{max}}{n} \leq 0$ almost surely (the probability is over the randomization in the forecaster if it is random). This means that for large $n$, the loss of the forecaster is essentially at least as small as that of any fixed strategy. As mentioned in the introduction, the problem addressed in this paper is of finding a sequential probability assignment $P^{(u)}_t(\cdot)$ such that the resulting prediction scheme \eqref{eq:70} is Hannan-consistent for a large class of loss functions. We will focus mainly on bounding the \textit{expected} loss (over the predictor's randomization), because it also leads to almost-sure bounds by applying the strong law of large numbers. The maximum \textit{expected} regret is defined as:
\begin{equation}\label{eq:109b}
\overline{\mathcal{R}}_\tsubs{max} = \max_{\vr x_1^n} \E \left[ \hat L_n - L^*_n \right]
,
\end{equation}

For some loss functions satisfying smoothness conditions \cite[Thm 3.1]{Nicolo}\cite[Thm 1]{FederMerhav93}, the forecasting strategy known as ``Follow the Leader'' (FL), which chooses at each time the best strategy in retrospect $\hat b_t^\tsubs{(FL)} = \argmin{b} \sum_{i=1}^{t-1} l(b, x_i)$, is Hannan consistent. Rewriting the above as $\hat b_t^\tsubs{(FL)} = \argmin{b} \sum_{x \in \mathcal{X}} \frac{N_{t-1}(x)}{t-1} l(b, x)$, it can be interpreted as an implementation of \eqref{eq:70} where the universal probability assignment equals the empirical frequencies $P^{(u)}_t(\cdot) = \frac{N_{t-1}(x)}{t-1}$. In other words, for this family of loss functions, there is a simple solution for $P^{(u)}_t(\cdot)$, namely the empirical distribution. However this class of loss functions where FL is universal, is rather limited.

For a probability assignment to be ``general'' enough, one would want to cover, at the least, the family of discrete-strategy, discrete-state loss functions, presented by Hannan \cite{Hannan57}. For this family, the loss function can be represented by a general $|\mathcal{B}| \times |\mathcal{X}|$ matrix specifying the loss for each strategy and each state of nature. It is well known \cite[\S4]{Nicolo} and straightforward to see that randomization is required in order to cover this class: consider the 0-1 loss case, i.e. binary sequences $\mathcal{X}=\mathcal{B}=\{0,1\}$ with $l(b,x) = \Ind(b \neq x)$, where the total loss is the number of errors. For this loss function, no deterministic predictor yields Hannan-consistency, because for each deterministic predictor there exists a sequence which fails the predictor completely, by choosing the next outcome as the opposite of the predictor's choice, while the loss of the best fixed predictor is at most $n/2$. Because a deterministic $P^{(u)}_t(\cdot)$ inevitably leads to a deterministic predictor \eqref{eq:70}, this implies a random $P^{(u)}_t(\cdot)$ is required, in general.

For the binary 0-1 loss problem, Feder, Merhav and Gutman \cite{Gutman} used a small dither when the empirical probability is close to $\half$, which effectively avoids a decision when the frequencies of $0,1$ are nearly equal.\footnote{It is interesting to note that for the 0-1 loss problem their forecaster is equivalent to a ``Follow the Perturbed Leader'' forecaster with a uniform distribution (see below) and also equivalent to the forecaster proposed here.} For this specific problem, the optimal solution (in the sense of minimax regret) is known exactly and was presented by Cover \cite{CoverPrague}. While the optimal dither in this problem is different than the straight line used by Feder, Merhav and Gutman, and is not known in general, this is of no consequence in the current problem, as we are only considering Hannan consistency. This solution, as well as the small bias from the empirical distribution which is required in the log-loss problem \eqref{eq:54}, motivates the following choice of $P^{(u)}_t(\cdot)$: add a small dither to $N_{t-1}(x)$ (the counts of events in the past) and re-normalize. As shown below, this solution achieves Hannan-consistency for any bounded loss function and for the log loss.

The proposed forecaster is reminiscent of the scheme termed ``Follow the Perturbed Leader'' (FPL), originally proposed by Hannan \cite{Hannan57}, in which the decision in obtained by adding a small dither to the accumulated loss of every reference strategy and then choosing the best one. Indeed, dithering the frequencies is similar, but not equivalent, to dithering the accumulated losses, and our proof technique for the bounded loss case borrows from Kalai and Vempala's \cite{KalaiV05}. Following this similarity we term the scheme proposed here ``Follow the Perturbed Frequency'' (FPF). Notice, however, that FPL is defined, in general, only when the number of strategies is finite, while FPF is defined, in general, only when the number of outcomes (states) is finite, and does not have to assume the number of strategies is finite. On the other hand, FPL can deal with more general forms of the problem, including time-varying loss functions.

The problem considered here is a close relative of the calibration problem \cite[\S4.5]{Nicolo}, i.e. the problem of estimating from an individual sequence, probability forecasts that pass certain consistency tests. The problems are related in that, in both cases it is shown possible to generate from empirical data collected from an individual sequence, probability assignments that appear to operate as well as forecasts which are based on knowledge of the ``true'' statistical model. Also, randomization is essential in both cases. However, none of the problems is a special case of the other: the probability assignment shown here is not necessarily calibrated, and a calibrated probability assignment does not necessarily satisfy the requirements of the current problem.\footnote{Consider for example the 0-1 loss problem, and a sequence containing an equal number of zeros and ones. Any probability forecaster yielding only values in the range $0.5 \pm \epsilon$ is $\epsilon$-calibrated, while the decisions based on these probabilities (when plugged into \eqref{eq:70}) can be arbitrary (depending on whether the probability is smaller or larger than $0.5$), and can yield arbitrarily bad (or good) aggregate losses.}

In this paper, in order to simplify matters, only fixed strategies are considered. As one of our motivations is to rationalize the behavior of learning probabilities from the past, it is enough to consider fixed strategies in order to see the advantage of this behavior. The extension to dynamic reference strategies is unfortunately not immediate as in the setting of prediction with expert advice \cite[\S2]{Nicolo}, where dynamic strategies can be turned into fixed ones by simple enumeration (i.e. replacing the strategy with the index of the strategy), because we explicitly assume a fixed loss function. However in some cases, the core of the prediction problem lies in competing with fixed strategies. For example, reference strategies defined by states (such as Markov predictors or finite state machines), can be considered as fixed strategies in each sub-sequence belonging to the same state.

%For general discrete loss functions, if there are two states $x, x'$, whose respective best strategies $b,b'$ are different,  $\max_x l(b_t, x)$ . ****
%$b_i$ $\sum_i \max_x l(b_i,x) > \sum

\section{Main results}\label{sec:main_results}
Let $N_{t}(x)$ be number of times a specific $x$ occurred in the sequence $\vr x$ up to and including time $t$. The universal sequential probability assignment is defined as:
\begin{equation}\begin{split}\label{eq:B109b}
P^{(u)}_t(x)
&=
c_t \cdot (N_{t-1}(x) + h_t \cdot u_t(x))
\\&=
\frac{N_{t-1}(x) + h_t \cdot u_t(x)}{t-1 + h_t \cdot \sum_{x' \in \mathcal{X}} u_t(x')}
\end{split}\end{equation}
where $c_t = \sum_{x \in \mathcal{X}} (N_{t-1}(x) + h_t \cdot u_t(x))$ is the normalizer guaranteeing unit sum. $u_t(x) \sim U[0,1]$ is a random dither which is assumed to be uniformly distributed, i.i.d. over different $x$ and $t$ (dependence over $t$ does not affect the expected regret). $h_t$ is a non-decreasing positive sequence. Our philosophical considerations (i.e. justifying probabilistic behavior) motivate keeping $h_t$ as general as possible rather than finding a specific optimal sequence $h_t$ for each problem.

The FPF predictor, for any loss function $l(b,x)$ is defined by:
\begin{equation}\label{eq:B114}
b^\tsubs{(FPF)}_t
=
\argmin{b \in \mathcal{B}} \underset{\vr X \sim P^{(u)}_t(x)}{\E} \left[ l(b, X) \right]
\end{equation}

\begin{theorem}\label{theorem:Pu_is_universal}
Assuming $h_t = h_1 \cdot t^{\alpha}$, with $\alpha \in (0,1)$, the FPF predictor is Hannan-consistent for any bounded loss function and for the log-loss. Therefore under these conditions, $P^{(u)}_t(x)$ defined in \eqref{eq:B109b} is a universal probability assignment for the class.
\end{theorem}

This theorem is based on the two following theorems:
\begin{theorem}\label{theorem:FPF_bounded_loss}
Assume the loss function is bounded $| l(b,x) | \leq R$. Then:
\begin{enumerate}
\item The expected regret of FPF is upper bounded by
\begin{equation}\label{eq:216}
\overline{\mathcal{R}}_{\max} \leq 2 R \sum_{t=1}^n h_t^{-1} +  2 R |\mathcal{X}| h_{n}
\end{equation}
\item Particularly, for any $h_t = h_1 \cdot t^{\alpha}$, with $\alpha \in (0,1)$, the normalized expected regret $\frac{1}{n} \overline{\mathcal{R}}_{\max}$ tends to zero with $n$.
\item For $h_t = \sqrt{\frac{2 t}{|\mathcal{X}|}}$, $\frac{1}{n} \overline{\mathcal{R}}_{\max} \leq 4 R \sqrt{\frac{2 |\mathcal{X}|}{n}}$.
\end{enumerate}
\end{theorem}

\begin{corollary_in_theorem}\label{corollary:FPF_bounded_loss_relaxation}
The theorem holds under a milder condition, that the loss function is bounded only for the set of optimizing strategies, defined as \begin{equation}\label{eq:215}
\mathcal{B}_\tsubs{opt} = \left\{ \argmin{b \in \mathcal{B}} \sum_{x \in \mathcal{X}} \lambda(x) l(b,x): \lambda(x) \geq 0, \exists x: \lambda(x)>0 \right\}
\end{equation}
and where $R = \sup_{x \in \mathcal{X}, b \in \mathcal{B}_\tsubs{opt}} l(b,x)$. Particularly, the theorem holds for the $L_2$ norm loss, $l(\vr b,\vr x) = \| \vr b - \vr x \|^2$ for $\mathcal{X} \subset \mathbb{R}^d$ ($|\mathcal{X}| < \infty$), and $\mathcal{B} =\mathbb{R}^d$. In that case $R = \max_{\vr x, \vr x' \in \mathcal{X}} \| \vr x - \vr x' \|^2$ is the squared diameter of the set $\mathcal{X}$.
\end{corollary_in_theorem}
Notice that in the most general case without any limitations (such as on magnitude), it is generally impossible to devise a universal scheme for the $L_2$ norm loss that beats the best fixed strategy, i.e. the empirical mean up to a constant, and it is made possible in the current problem by the assumption that $\mathcal{X}$ is finite.

The proof of Theorem~\ref{theorem:FPF_bounded_loss} is similar in spirit to the proof of Kalai and Vempala \cite{KalaiV05} for the FPL forecaster, as the perturbation on $N_t(x)$ can be translated to a perturbation on the accumulated loss.

\begin{theorem}\label{theorem:FPF_log_loss}
For the case of the log-loss, where $b(x)$ is a probability distribution over $\mathcal{X}$ and $l(b,x)= \log \left( \frac{1}{b(x)} \right)$,\footnote{All $\log$-s in this paper are in the natural base.} the expected regret of FPF satisfies:
\begin{enumerate}
\item
\begin{equation}\label{eq:230}
\overline{\mathcal{R}}_{\max} \leq \sum_{t=1}^n \frac{|\mathcal{X}| h_t - 1}{t}
+
\sum_{t=1}^n \frac{1}{\lfloor \frac{t-1}{|\mathcal{X}|} \rfloor + h_t}
\end{equation}
\item Particularly, for any $h_t = h_1 \cdot t^{\alpha}$, with $\alpha \in [0,1)$, the normalized expected regret $\frac{1}{n} \overline{\mathcal{R}}_{\max}$ tends to zero with $n$.
\item For constant $h_t$ the expected regret behaves like $O(\log n)$ and specifically for the choice $h = |\mathcal{X}|^{-1}$, $\overline{\mathcal{R}}_{\max} \leq |\mathcal{X}| \log (n)$.
\end{enumerate}
\end{theorem}
Regarding the last case, notice that this redundancy is similar to the redundancy obtained with Laplace's estimator and approximately twice the redundancy obtained using Kritchevsky-Trofimov's (which is approximately $\frac{|\mathcal{X}|-1}{2} \log n$). However notice that the target of the FPF forecaster was not to produce optimal redundancy for specific loss functions.

The proofs of the theorems stated above appear in \onlyfull{Section~\ref{sec:proofs} below.}\onlyconf{the full paper \cite{YL_FPF}.}

\section{Implications on the understanding of probability}\label{sec:philo_part}

\subsection{Initial probabilities}\label{sec:philo_initial_probabilities}
A basic question in the application and philosophy of probability theory is: where do initial probabilities originate from (see, e.g.  \cite{Weatherford_ProbabilityFoundations}) ? The fact is, that in many situations a probability distribution is deduced from the relative frequency of events in the past. While this deduction may be justified based on some stationarity assumption, it is often used exactly in those situations where precise analysis of the source of events is not possible, and therefore the assumption that the frequency of events in the future would be similar to their frequency in the past is not necessarily justified. In spite of this, we often deduce a probability distribution based on past statistics and use this probability for decision making with regards to future events. It seems that not only humans but also animals use this principle \cite{ReznikovaAnts2012}.

One motivation for the problem posed in Section~\ref{sec:problem_statement}, of searching for a universal probability assignment, is the attempt to justify this behavior based on mathematical, rather than physical assumptions. The theory of universal prediction of individual sequences, or repetitive games, seems a good framework for this purpose, because it facilitates deduction from the past, without assumptions that the past indicates anything with respect to the future. The existing universal prediction schemes are less suitable for this purpose since they determine the next strategy in a contrived way, as a function of the past frequencies and the loss function, whereas in the probability-based decision making, it is assumed that there exist a single ``true'' probability.

The success of the FPF predictor for a large set of loss functions, indicates that indeed it is useful to rely on past frequencies, and draw from them a ``probability'' distribution, even if the future is arbitrary. The dither may be interpreted as the assumption that the future would be similar but not identical to the past, and prevents using a too ``decisive'' strategy (such as choosing '0' or '1' in the 0-1 loss case), based on a small change in the frequencies. It would be farfetched to claim that this is \emph{the} justification for using probabilities: clearly the reason is related to the regularity that many natural processes exhibit; however it supplements our intuitive understanding by showing that even if these assumptions fail, there is still benefit in learning probabilities from the past.

%(about the 0-1 loss) It is interesting to think how this aligns with our intuition about learning from the past. Suppose that our evaluation of the probability of a coin to fall on tail is close to $\half$ (e.g. 0.49 or 0.51). Under these conditions we would usually refrain from betting on the coin's value, because the average gain from such a bet is small while the potential loss if our assumptions are wrong, is high. If we are forced to bet (as in the 01 problem) then betting randomly seems a good idea.
%This prevents a decision when the odds are equal (so to speak). This seems to parallel what we do in practice: we never exacly trust our probabilistic forecasts.
%

\subsection{Meaning of probability}\label{sec:philo_meaning_of_p}
In the previous section we tried to justify a specific choice of a probability. However, probability itself is not a well defined concept, and many attempts to explain or justify its use have been made. A good introduction to these philosophical questions can be found in \cite{Weatherford_ProbabilityFoundations} (for a quick overview see \cite[Chap.~\ref{PhD-chap:epilog}]{YL_PhdThesis}). While there is no dispute on the mathematical axiomatic theory dealing with probability functions, the meaning of probability, and the justification for using it are questionable.

In a nutshell, the main interpretations to probability are the relative frequency approach, a-priori or logical approach an the subjectivistic approach. Relative frequency theories interpret probability as the limiting frequencies in very large groups of events (called ``collectives''). A-priori theories interpret probability as logical relation between sentences, and an extension of formal logic: the attributes ``true'' and ``false'' are represented by probabilities of 1 and 0, and are extended by adding a range of probabilities in between. Subjectivistic theories interpret probability as a measure of the degree of belief of a certain person in a certain proposition, and therefore its value is not unique.

A main issue in all interpretations is what probability means with respect to the future. The current results can be interpreted under the framework of the subjectivistic theories, which view probability as a tool for decision making, i.e. probability is just the relative weight that we put on each future event when making decisions. Because under subjectivistic theories any probability is valid, there is a problem of justifying any specific choice of a probability assignment, as well as the merit of making decisions according to probabilistic considerations.

The current results can be thought of as a partial resolution to this question: the suggestion of learning probability from the past by biasing or dithering past frequencies, is a good one in the sense that it is better than any fixed behavior (and as a result, of making decisions according to any fixed probability). This demonstrates a clear merit in following probabilistic considerations, which is not dependent on any assumptions with respect to the real world (the process $\vr x_t$).

There are some issues, however, with this interpretation. First, the problem setting is limited, compared to our actual use of probability. Learning from experience extends far beyond the framework of repetitive games and constant loss functions, as we usually deduce probabilities from the past and use them to solve \emph{new} problems. Also the fact $\mathcal{X}$ is assumed discrete is somewhat limiting, although it may be sufficient to justify probabilistic intuition, which is fundamentally based on distributions on finite sets (such as coins and dice).

But the main weakness of this interpretation is that it relies on randomness for generating the universal probability assignment $P^{(u)}$ (and as a result, the claims we can make are also probabilistic), and so it may lead to a cyclic argument of explaining probability by using probability. The randomness used here is in a restricted form of ``controlled randomness'' which is generated by the forecaster. I.e. if we believe it is possible to draw random coins, it is enough for this interpretation to hold and be meaningful. An alternative assumption is pseudo-randomness, i.e. assume that we can generate the dither not randomly, but such that ``nature'' (drawing the next $x_t$) cannot guess it, and it appears effectively random. Unfortunately, like in many other theories, we are not able to escape some form of ``belief'' or conjecture with respect to the future.

Another way to avoid the need for randomness is to avoid problems such as the 0-1 loss case, in which one is forced to bet, problems that are insolvable without randomness. For example, if the loss is convex with respect to the strategy, then the loss when taking the expected value of a random strategy $b$ is always better than the expected loss when $b$ is random. In this case, the forecaster can make a deterministic decision: replace \eqref{eq:B114} with $\hat b_t = \E \left\{ \hat b^\tsubs{(FPF)}_t \right\}$, where the expected value is with respect to the randomness of $P^{(u)}$. This can be thought of as a different rule for making decisions based on the past: take as probability the empirical frequencies in the past, however when making a decision which changes significantly with respect to small variations in the probability, take the average decision over these small variations. This rule is deterministic and aligns with intuition, however the restriction to ``smooth'' loss functions may be too limiting.

Another question that would naturally arise with respect to this explanation is how it aligns with the fact that, at least in the theoretical application of probability theory (e.g. estimation theory, communication theory) we do not use dithers in our probabilities. It seems that the idea of dithering the probabilities is a similar notion to the idea of checking sensitivity of a given solution to the probabilistic assumptions. In case the solution to a given problem does not depend crucially on the exact probability values, adding the dither is indeed redundant. On the other hand, if the solution depends crucially on a small change in the probabilistic assumptions, it may be reasonable to doubt its operation in the real world.

\onlyfull{
\section{Proofs}\label{sec:proofs}
\subsection{Proof of Theorem~\ref{theorem:FPF_bounded_loss}}
The proof follows the same line of thought of Kalai-Vemplala \cite{KalaiV05}: first, the regret of a clairvoyant forecaster using $x_t$ in addition to $\vr x_1^{t-1}$ is bounded. Then, the difference in performance between the clairvoyant forecaster and the proposed forecaster is bounded, by using the fact that some of the dither works in the same direction as the the difference between them.

\subsubsection{Definitions}
The cumulative loss for playing the constant strategy $b$ up to time $t$ is $L_t(b) = \sum_{i=1}^t l(b, x_t)$. We denote for brevity $\bbest\{L(b)\} \defeq \argmin{b} L(b)$ the best strategy for cumulative loss function $L(b)$.

The optimal fixed (a-posteriori) best fixed strategy is $\bbest\{L_n(b)\}$ and has loss $L_n^* = L_n(\bbest\{L_n(b)\})$. As another example to clarify the notation, the FL predictor can be written as $\bbest\{L_{t-1}(b)\}$, and the FPL predictor \cite{KalaiV05} can be written $\bbest\{L_{t-1}(b) + p_t(b)\}$ where $p_t(b)$ is a random perturbation.

Let us define the dithered count at time $t-1$ as
\begin{equation}\label{eq:368a}
\Nfpf_{t-1}(x) \defeq N_{t-1}(x) + h_{t} u_{t}(x)
,
\end{equation}
and the respective dithered accumulated loss as
\begin{equation}\label{eq:372}
\Lfpf_{t-1}(b) \defeq \sum_x l(b,x) \Nfpf_{t-1}(x)
.
\end{equation}
This loss could be thought of as the loss during a sequence which is an extension of the actual sequence with some random states. Notice the distinction between $\hat L_n$ defined in \eqref{eq:96}, which is the loss of the universal predictor, and $\Lfpf_t$ which is the accumulated loss whose minimization yields the predictor. The distribution $P^{(u)}$ is proportional to $\Nfpf_{t-1}(x)$, and thus the FPF forecaster is equivalent to optimizing the dithered loss:
\begin{equation}\begin{split}\label{eq:B114x}
b^\tsubs{(FPF)}_t
&=
\argmin{b \in \mathcal{B}} \underset{\vr X \sim P^{(u)}_t(x)}{\E} \left[ l(b, X) \right]
\\&=
\argmin{b \in \mathcal{B}} \sum_x P^{(u)}_t(x) l(b, x)
\\&=
\argmin{b \in \mathcal{B}} \left[ c_t \cdot \sum_x \Nfpf_{t-1}(x) l(b, x) \right]
\\&=
\bbest\{\Lfpf_{t-1}(b)\}
.
\end{split}\end{equation}
Notice that the constant $c_t$ does not affect the minimum.

\subsubsection{Bounding the expected loss}
In terms of the expected loss $\E \left[ \hat L_n \right] =  \sum_{t=1}^n \E \left[ l(\hat b_t, x_t) \right]$, only the marginal distribution of $\hat b_t$ matters, and therefore dependence between $u_t(x)$ at different times does not affect the expected loss. Therefore in this section, we assume all $u_t$ are equal, $u_t(x) = u_1(x)$.

We start by analyzing a clairvoyant predictor which includes also the state $x_t$ into the prediction. For this purpose, let us define analogously to \eqref{eq:368a}-\eqref{eq:372}:
\begin{equation}\label{eq:392}
\Ncl_t(x) \defeq N_t(x) + h_{t} u_{t}(x), \qquad \Lcl_t(b) \defeq \sum_x l(b,x) \Ncl_t(x)
\end{equation}
where for $t=0$ we define $h_0 = 0$, and note that $N_0(x)=0$ by definition, and therefore $\Ncl_0(x) = 0$ and $\Lcl_0(b) = 0$. We consider the loss of the predictor $\hat b_t = \bbest\{\Lcl_t(b)\}$:
\begin{equation}\begin{split}\label{eq:383}
& \sum_{t=1}^n l(\bbest\{\Lcl_{t}(b)\}, x_t)
\\ & \stackrel{(a)}{=}
\sum_{t=1}^n \sum_x \left[ l(\bbest\{\Lcl_{t}(b)\}, x) (N_t(x) - N_{t-1}(x)) \right]
\\&=
\sum_{t=1}^n \sum_x \left[ l(\bbest\{\Lcl_{t}(b)\}, x) (\Ncl_t(x) - \Ncl_{t-1}(x)) \right]
\\& \qquad -
\sum_{t=1}^n \sum_x \left[ l(\bbest\{\Lcl_{t}(b)\}, x) (h_{t} u_{t}(x) - h_{t-1} u_{t-1}(x)) \right]
\end{split}\end{equation}
where in (a) we used $N_t(x) - N_{t-1}(x)$ as an indicator function $\Ind(x_t = x)$.
The first part can be bounded as:
\begin{equation}\begin{split}\label{eq:395}
&
\sum_{t=1}^n \sum_x \left[ l(\bbest\{\Lcl_{t}(b)\}, x) (\Ncl_t(x) - \Ncl_{t-1}(x)) \right]
\\& =
\sum_{t=1}^n \left[ \Lcl_t(\bbest\{\Lcl_{t}(b)\}) -  \Lcl_{t-1}(\bbest\{\Lcl_{t}(b)\}) \right]
\\& \stackrel{(a)}{\leq}
\sum_{t=1}^n \left[ \Lcl_t(\bbest\{\Lcl_{t}(b)\}) -  \Lcl_{t-1}(\bbest\{\Lcl_{t-1}(b)\}) \right]
\\& \stackrel{(b)}{=}
\Lcl_n(\bbest\{\Lcl_{n}(b)\}) - \Lcl_{0}(\bbest\{\Lcl_{0}(b)\})
\\& =
\Lcl_n(\bbest\{\Lcl_{n}(b)\})
\\& \leq
\Lcl_n(\bbest\{L_{n}(b)\})
\\&=
L_n(\bbest\{L_{n}(b)\}) + h_{n} \sum_x l(\bbest\{L_{n}(b)\},x) u_{n}(x)
\\&\leq
L_n(\bbest\{L_{n}(b)\}) +  R |\mathcal{X}| h_{n}
\\&=
L^*_n +  R |\mathcal{X}| h_{n}
,
\end{split}\end{equation}
where we used (a) the fact that $\bbest\{\Lcl_{t-1}(b)\}$ is optimized for $\Lcl_{t-1}$ and (b) the sum of the telescopic series.
For the second sum in \eqref{eq:383}, let us use the assumption $u_t = u_1$. Then:
\begin{equation}\begin{split}\label{eq:428}
&
\left| \sum_{t=1}^n \sum_x \left[ l(\bbest\{\Lcl_{t}(b)\}, x) (h_{t} u_{t}(x) - h_{t-1} u_{t-1}(x)) \right] \right|
\\&=
\left| \sum_{t=1}^n \sum_x \left[ l(\bbest\{\Lcl_{t}(b)\}, x) (h_{t}  - h_{t-1}) u_{1}(x) \right] \right|
\\&\leq
\sum_{t=1}^n \sum_x R |h_{t}  - h_{t-1}|
\\&\stackrel{(a)}{=}
|\mathcal{X}| R  \sum_{t=1}^n (h_{t}  - h_{t-1})
\\& \stackrel{(b)}{=}
R |\mathcal{X}| h_{n}
,
\end{split}\end{equation}
where we  used (a) the assumption that the sequence $h_t$ is non decreasing, and (b) the definition $h_0 = 0$. Combining \eqref{eq:428} and \eqref{eq:395} into \eqref{eq:383} yields:
\begin{equation}\label{eq:444}
\sum_{t=1}^n l(\bbest\{\Lcl_{t}(b)\}, x_t)
\leq
L^*_n +  2 R |\mathcal{X}| h_{n}
\end{equation}

The next step is to bound the performance difference between the clairvoyant predictor $\bbest\{\Lcl_{t}(b)\}$ and the FPF forecaster $b^\tsubs{(FPF)}_t=\bbest\{\Lfpf_{t-1}(b)\}$. The key is that the new element added to $\Lcl_t(b)$ is $l(b,x_t)$, and the dither element $u(x_t)$ (i.e. belonging to the state that actually happened at time $t$) contributes an offset in the same direction, which cancels this addition or most values of $u(x_t)$. For this purpose let us write the accumulated losses as:
\begin{equation}\begin{split}\label{eq:454}
\Lfpf_{t-1}(b)
&=
\sum_x l(b,x) (N_{t-1}(x) + h_{t} u_{t}(x))
\\& =
L_c + h_t u(x_t) l(b,x_t)
\\
\Lcl_t(b) &= \sum_x l(b,x) (N_t(x) + h_{t} u_{t}(x))
\\&=
\Lfpf_{t-1}(b) + l(b,x_t)
\\&=
L_c + (h_t u(x_t) + 1) l(b,x_t)
\end{split}\end{equation}
where we defined
\begin{equation}\label{eq:459}
L_c = \sum_x l(b,x) N_{t-1}(x) + \sum_{x \neq x_t} l(b,x) h_{t} u_{t}(x)
.
\end{equation}
Noticing that the common part $L_c$ is independent of $u(x_t)$, we compute the conditional expectation given $L_c$ for each of the predictors:
\begin{equation}\begin{split}\label{eq:B178b1}
& \E \left[  l(\bbest\{\Lfpf_{t-1}(b)\}, x_t) \Big| L_c \right]
\\&=
\E \left[  l(\bbest\{L_c + h_t u(x_t) l(b,x_t) \}, x_t)  \Big| L_c \right]
\\&=
\int_{v=0}^1 l(\bbest\{L_c + h_t l(b,x_t) v \}, x_t) dv
\\&=
\int_{v=0}^1 g(v) dv
,
\end{split}\end{equation}
where we defined for brevity $g(v) = l(\bbest\{L_c + h_t l(b,x_t) v \}, x_t)$, and
\begin{equation}\begin{split}\label{eq:B178b2}
& \E \left[  l(\bbest\{\Lcl_{t}(b)\}, x_t) \Big| L_c \right]
\\&=
\E \left[   l(\bbest\{L_c + (h_t u(x_t) + 1) l(b,x_t)\}, x_t) \Big| L_c \right]
\\&=
\int_{v=0}^1   l(\bbest\{L_c + (h_t v + 1) l(b,x_t)\}, x_t)  dv
\\&=
\int_{v=0}^1   l(\bbest\{L_c + h_t  (v + h_t^{-1}) l(b,x_t)\}, x_t)  dv
\\&=
\int_{v=h_t^{-1}}^{1+ h_t^{-1}} l(\bbest\{L_c + h_t l(b,x_t) v \}, x_t) dv
\\&=
\int_{v=h_t^{-1}}^{1+ h_t^{-1}} g(v) dv
\end{split}\end{equation}
The integrands in \eqref{eq:B178b1},\eqref{eq:B178b2} are equal. Let us temporarily assume that for $t \geq 1$, $h_t > 1$, so that the integration regions partially overlap. For most of the integration region, because the integrands are the same (no matter what $l(\cdot, x_t)$ evaluates to), and the integration regions overlap, they cancel out, and we remain with the contribution of the edges where there is no overlap:
\begin{equation}\begin{split}\label{eq:B178b}
& \E \left[  l(\bbest\{\Lfpf_{t-1}(b)\}, x_t) - l(\bbest\{\Lcl_{t}(b)\}, x_t) \Big| L_c \right]
\\& \stackrel{\eqref{eq:B178b1},\eqref{eq:B178b2}}{=}
\int_{v=0}^1 g(v) dv - \int_{v=h_t^{-1}}^{1+ h_t^{-1}} g(v) dv
\\& =
\int_{0}^{h_t^{-1}} g(v) dv - \int_{1}^{1+ h_t^{-1}} g(v) dv
\\&\leq
\int_{[0,h_t^{-1}] \cup [1,1+ h_t^{-1}]} | g(v) | dv
\\&\leq
2 \cdot R \cdot h_t^{-1}
.
\end{split}\end{equation}
Recall that we assumed $h_t \geq 1$. For $h_t \leq 1$ the bound \eqref{eq:B178b} is trivially true (because the RHS is at least $2R$), and therefore it holds for all $h_t$.

Applying the iterated expectations law and accumulating \eqref{eq:B178b} yields:
\begin{equation}\label{eq:B178c}
\E \left[ \sum_{t=1}^n l(\bbest\{\Lfpf_{t-1}(b)\}, x_t) - \sum_{t=1}^n l(\bbest\{\Lcl_{t}(b)\}, x_t) \right] \leq 2 R \sum_{t=1}^n h_t^{-1}
\end{equation}

which, together with \eqref{eq:444} yields:
\begin{equation}\begin{split}\label{eq:B178d}
&
\E \left[ \sum_{t=1}^n l(\bbest\{\Lfpf_{t-1}(b)\}, x_t)  \right] - L^*_n
\\&=
\E \left[ \sum_{t=1}^n l(\bbest\{\Lfpf_{t-1}(b)\}, x_t) - \sum_{t=1}^n l(\bbest\{\Lcl_{t}(b)\}, x_t) \right]
\\& \qquad +
\E \left[ \sum_{t=1}^n l(\bbest\{\Lcl_{t}(b)\}, x_t) \right] - L^*_n
\\& \stackrel{\eqref{eq:B178c},\eqref{eq:444}}{\leq}
2 R \left( \sum_{t=1}^n h_t^{-1} +  |\mathcal{X}| h_{n} \right)
\defeq
\Delta
,
\end{split}\end{equation}
which proves the first claim of the theorem.

\subsubsection{Choices of the dither amplitude sequence}
There remains the question of selecting the sequence of dither amplitudes $h_t$.
For a given horizon $n$, a simple calculation shows, that the best choice in terms of minimizing $\Delta$ is a constant $h_t$, which equals $\sqrt{\frac{n}{|\mathcal{X}|}}$. As will be seen below, a good choice of a varying $h_t$ that yields an infinite horizon solution (i.e. in which the setting of $h_t$ does not depend on the horizon $n$) is $h_t = \sqrt{\frac{2 t}{|\mathcal{X}|}}$. However, because in real life we do not choose an ``optimal'' $h_t$, it is first desired to show that for a wide range of choices, the resulting predictor's expected regret tends to zero. This analysis is rather straightforward and reoccurs in many developments of this kind \cite[Ex 4.7]{Nicolo}\cite{KalaiV05}. So, let us choose $h_t = h_1 \cdot t^{\alpha}$, with $\alpha \in (0,1)$. Then,
\begin{equation}\begin{split}\label{eq:B246}
\sum_{t=1}^n h_t^{-1}
&=
h_1^{-1} \sum_{t=1}^n t^{-\alpha}
\leq
h_1^{-1} \left(1 + \int_{x=1}^{n} x^{-\alpha} dx  \right)
\\&=
h_1^{-1} \left(1 + \frac{1}{1-\alpha} (n^{1-\alpha} - 1) \right)
\leq
h_1^{-1} \frac{1}{1-\alpha} n^{1-\alpha}
\end{split}\end{equation}

Substituting in \eqref{eq:B178d} yields:
\begin{equation}\begin{split}\label{eq:B271}
\frac{\Delta}{n}
&=
\frac{2 R}{n} \left( \sum_{t=1}^n h_t^{-1} + |\mathcal{X}| \cdot h_n \right)
\\& \leq
2 R \left( h_1^{-1} \frac{1}{1-\alpha} n^{-\alpha} + |\mathcal{X}| \cdot h_1 n^{\alpha-1} \right)
\end{split}\end{equation}
For any $\alpha \in (0,1)$ and any $h_1$, this yields $\frac{\Delta}{n} \arrowexpl{n  \to \infty} 0$, i.e. Hannan's consistency. It is straightforward to see that the best choice is obtained by $\alpha=\half$ and $h_1 = \sqrt{\frac{2}{|\mathcal{X}|}}$, which yields:
\begin{equation}\label{eq:B271b}
\frac{\Delta}{n}
=
\frac{2 R}{\sqrt{n}} \left( 2 h_1^{-1} + |\mathcal{X}| \cdot h_1 \right)
=
4 R \sqrt{\frac{2 |\mathcal{X}|}{n}}
\end{equation}

\subsubsection{Proof of Corollary~\ref{corollary:FPF_bounded_loss_relaxation}}
To prove the corollary is it sufficient to notice that all strategies for which the loss is computed in the proof of Theorem~\ref{theorem:FPF_bounded_loss}, are in the aforementioned set of optimizing strategies. For the $L_2$ loss it is easy to see that the set of optimizing strategies is the convex hull of $\mathcal{X}$ (the strategy for given $\lambda(x)$ can be interpreted as a center of mass of $\mathcal{X}$ with varying weights to the different points).

\subsection{Proof of Theorem~\ref{theorem:FPF_log_loss} (Log loss)}\label{sec:proof_log_loss}
The sequence is $x_t$, $t=1,\ldots,n$. The accumulated loss for probability $q_t$ is $\sum_{t=1}^n \log \frac{1}{q_t(x_t)}$. The best fixed $q_t$ in hindsight is $q_t(x) = \hat P_{\vr x}(x)$ and yields $L^*_n = n \sum_{x} \hat P_{\vr x}(x) \log \frac{1}{\hat P_{\vr x}(x)}$. For the universal estimator proposed: $P^{(u)}_t(x) = c_t^{-1} \cdot (N_{t-1}(x) + h_t u_t(x))$, and it is easy to see that given $P^{(u)}_t(x)$, the choice of $q_t(\cdot)$, the probability distribution for the next state is just $q_t(\cdot) = \argmin{q} \underset{P^{(u)}_t(x)}{\E} \log \frac{1}{q(X)} = P^{(u)}_t(x)$

\begin{equation}\begin{split}\label{eq:342}
\E [\hat L_n]
&=
\E \sum_{t=1}^n \log \frac{1}{P^{(u)}_t(x_t)}
\\&=
s\sum_{t=1}^n \left[ \E \log (c_t) - \E \log (N_{t-1}(x_t) + h_t u_t(x_t))  \right]
\end{split}\end{equation}

In general, in order to achieve a small regret for the $\log$ loss, it is required that the overall contribution of $P^{(u)}_t(x_t)$ for all occurrences of a certain state $x_t = x$, would approximate $\hat P_{\vr x}(x)$. However the most important property, which is not satisfied by FL, is not to give a probability too close to $0$ for a certain state $x$ on its first appearance in the sequence. I.e. if $N_{t-1}(x_t) = 0$, it is required that $\E \log (N_{t-1}(x_t) + h_t u_t(x_t)) = \E \log (h_t u_t(x_t))$ is finite. Indeed it is easy to verify that this holds.

Following is the detailed calculation and bounding for $\E [\hat L_n]$. The normalized $c_t$ is bounded as:

\begin{equation}\label{eq:346}
c_t = \sum_x (N_{t-1}(x_t) + h_t u_t(x)) \leq t-1 + |\mathcal{X}| h_t
\end{equation}

In the below, denote for conciseness $N_{t-1}(x_t) = v$
\begin{equation}\label{eq:368}
  \begin{split}
&
\E \log (N_{t-1}(x_t) + h_t u_t(x_t))
\\&=
\E \log (v + h_t u_t(x_t))
=
\int_{0}^1 \log (v + h_t y) dy
\\&=
h_t^{-1} \int_{v}^{v + h_t} \log (y) dy
\\&=
h_t^{-1} \left[ y \log y - y \right]_{v}^{v + h_t}
\\&=
h_t^{-1} \left[ (v + h_t) \log (v + h_t) - v \log v - h_t \right]
\\& \stackrel{(a)}{=}
h_t^{-1} \left[ h_t \log (v + h_t) + v \log \left(1 + \tfrac{h_t}{v}\right) - h_t \right]
\\& \stackrel{\log x \geq \frac{x}{1+x}}{\geq}
h_t^{-1} \left[ h_t \log (v + h_t) + v \frac{\tfrac{h_t}{v}}{1+ \tfrac{h_t}{v}}- h_t \right]
\\&=
\log (v + h_t) - \frac{h_t}{v+ h_t}
\\&=
\log (N_{t-1}(x_t) + h_t) - \frac{h_t}{N_{t-1}(x_t)+ h_t}
.
\end{split}
\end{equation}
In (a) notice that $v=0$ is a special case. using $0\log0=0$, it is easy to verify that in this case the the expression before (a) for $v=0$ equals $\log (h_t) - 1$, and therefore the inequality holds.

Returning to \eqref{eq:342}:

\begin{equation}\begin{split}\label{eq:342b}
\E [\hat L_n]
&=
\sum_{t=1}^n \left[ \E \log (c_t) - \E \log (N_{t-1}(x_t) + h_t u_t(x))  \right]
\\&\leq
\sum_{t=1}^n \log \left(t-1 + |\mathcal{X}| h_t  \right)
\\& \qquad + \sum_{t=1}^n \left[  - \log (N_{t-1}(x_t) + h_t) + \frac{h_t}{N_{t-1}(x_t)+ h_t}  \right]
\\&=
\sum_{t=1}^n \log \left(t-1 + |\mathcal{X}| h_t  \right)
\\& \qquad +
\sum_{x \in \mathcal{X}} \sum_{t: x_t = x} \left[ - \log (N_{t-1}(x) + h_t) + \frac{h_t}{N_{t-1}(x)+ h_t}  \right]
\end{split}\end{equation}

For the second sum, which was broken into the subsequenes in which a specific state $x$ appears, notice that $N_{t-1}(x)$ increases by $1$ between consecutive elements of the internal sum. The final value of $N_{t-1}(x)$ in the last element equals the total number of apperances of $x$, $N_n(x)$. At this point, it is beneficial to write $L^*_n$ in a similar form:
\begin{equation}\begin{split}\label{eq:394}
L^*_n
&=
n \sum_{x} \hat P_{\vr x}(x) \log \frac{1}{\hat P_{\vr x}(x)}
\\&=
\sum_{x} N_n(x) \log \frac{n}{N_{n}(x)}
\\&=
n \log n - \sum_{x} N_n(x) \log N_{n}(x)
\end{split}\end{equation}

A consequence of the bound on the size of a type class $|\mathcal{T}_P| \leq \exp(n H(P))$ \cite[Lemma II.2]{MethodOfTypes} is (considering the type defined by the sequence $x$, $\left( \ldots,\frac{N_n(x)}{n},\ldots \right)$ and taking the $\log$ of both sides):
\begin{equation}\begin{split}\label{eq:457}
\log \left( \frac{n!}{\prod_{x \in \mathcal{X}} N_n(x)!} \right)
& \leq
n \sum_{x \in \mathcal{X}} \frac{N_n(x)}{n} \log \frac{n}{N_n(x)}
\\& =
n \log n - \sum_{x \in \mathcal{X}} N_n(x) \log N_n(x)
\end{split}\end{equation}

Plugging into \eqref{eq:394} yields:
\begin{equation}\begin{split}\label{eq:394a}
L^*_n
& \geq
\log \left( \frac{n!}{\prod_{x \in \mathcal{X}} N_n(x)!} \right)
\\& =
\sum_{t=1}^n \log (t) - \sum_{x \in {\mathcal{X}}} \sum_{m=1}^{N_n(x)} \log (m)
\end{split}\end{equation}

Notice that this way of bounding $L^*_n$ is slightly non standard: rather than writing $L^*_n$ in a similar form to $L_U$, it would generally be simpler to write the bound on $\E [\hat L_n]$ using factorials, and simplify it using Stirling's approximation, obtaining a form similar to \eqref{eq:394}, however this approach does not hold for varying $h_t$.

Let us now assume $h_t$ is non-decreasing. Combining \eqref{eq:342b} with \eqref{eq:394a} yields:
\begin{equation}\begin{split}\label{eq:342c}
&
\E [\hat L_n] - L^*_n
\\&\leq
\sum_{t=1}^n \left( \log \left(t-1 + |\mathcal{X}| h_t  \right) - \log (t) \right)
\\& \qquad +
\sum_{x \in \mathcal{X}} \sum_{t: x_t = x} \Big[ \log(N_{t-1}(x) + 1)
\\& \qquad \qquad
- \log (N_{t-1}(x) + h_t) + \frac{h_t}{N_{t-1}(x)+ h_t} \Big]
\\& =
\sum_{t=1}^n \left( \log \left(1 + \frac{|\mathcal{X}| h_t - 1}{t}  \right) \right)
\\& \qquad +
\sum_{x \in \mathcal{X}} \sum_{t: x_t = x} \Big[ \log \left( 1+ \frac{1 - h_t}{N_{t-1}(x) + h_t} \right)
\\& \qquad + \frac{h_t}{N_{t-1}(x)+ h_t} \Big]
\\& \leq
\sum_{t=1}^n \frac{|\mathcal{X}| h_t - 1}{t}
\\& \qquad +
\sum_{x \in \mathcal{X}} \sum_{t: x_t = x} \left[ \frac{1 - h_t}{N_{t-1}(x) + h_t} + \frac{h_t}{N_{t-1}(x)+ h_t} \right]
\\&=
\sum_{t=1}^n \frac{|\mathcal{X}| h_t - 1}{t}
+
\sum_{x \in \mathcal{X}} \sum_{t: x_t = x}  \frac{1}{N_{t-1}(x) + h_t}
\\&=
\sum_{t=1}^n \frac{|\mathcal{X}| h_t - 1}{t}
+
\sum_{t=1}^n \frac{1}{N_{t-1}(x_t) + h_t}
\end{split}\end{equation}
Let us consider the sequence $\vr x$ that maximizes the second sum. As clear intuitively, and will be proven below, this sequence selects all states of $x$ in a round-robin fashion, which minimizes the growth rate of $N_{t-1}(x_t)$ and for which $N_{t-1}(x_t) = \lfloor \frac{t-1}{|\mathcal{X}|} \rfloor$.

First, for a given type (i.e. for given $\{N_n(x)\}_{x \in \mathcal{X}}$), consider the order that would yield the maximum. It is clear that the $m$-th occurrences of different states (assuming these states indeed occur at least $m$ times) should occur at consecutive $t$-s. In other words, the sequence $N_{t-1}(x_t)$ is non decreasing. Suppose that the opposite occurs, i.e. $N_{t-2}(x_{t-1}) > N_{t-1}(x_{t})$, then obviously $x_t \neq x_{t-1}$. Let us flip the order of these states, i.e. let $x_t' = x_{t-1}$ and $x_{t-1}' = x_t$, then as a result the counts will also flip, $N_{t-2}'(x_{t-1}') = N_{t-1}(x_{t})$ and $N_{t-1}'(x_{t}') = N_{t-2}(x_{t-1})$. This is easiest to see via an example: suppose the sequence is $\vr x = (c,a,a,b,c,c)$, then the counts $N_{t-1}(x_{t})$ are $0,0,\mathbf{1,0},1,2$. After flipping the states $t=3,4$ the sequence is $\vr x' = (c,a,b,a,c,c)$ and the counts are $0,0,\mathbf{0,1},1,2$. The elements pertaining to other times are not affected by this flip, while the sum of the two elements is now:
\begin{equation}\begin{split}\label{eq:537}
&
\frac{1}{N_{t-2}'(x_{t-1}') + h_{t-1}} + \frac{1}{N_{t-1}'(x_{t}') + h_{t}}
\\&=
\frac{1}{N_{t-1}(x_{t})  + h_{t-1}} + \frac{1}{N_{t-2}(x_{t-1}) + h_{t}}
\\&>
\frac{1}{N_{t-2}(x_{t-1})  + h_{t-1}} + \frac{1}{N_{t-1}(x_{t}) + h_{t}}
\end{split}\end{equation}
where the inequality holds because $h_{t} \geq h_{t-1}$ and $N_{t-2}(x_{t-1}) > N_{t-1}(x_{t})$. This can be seen by direct algebraic manipulation, or by using Lemma~\ref{lemma:hybrid_sums} with the convex function $f(x) = \frac{1}{x}$:

\begin{lemma}\label{lemma:hybrid_sums}
Let $f$ be a convex-$\cup$ function and $a_0, a_1, b_0, b_1$ satisfy $a_1 \geq a_0$ and $b_1 \geq b_0$, then
\begin{equation}\label{eq:536}
f(a_0+b_0) + f(a_1+b_1) \geq f(a_0 + b_1) + f(a_1 +b_0)
\end{equation}
I.e. the maximum sum is obtained by joining the smaller and bigger elements together.
\end{lemma}

\textit{Proof:}
Let us assume there is strict inequality at least in one of the pairs, otherwise the result holds trivially. Write the hybrid sums as convex combinations of the homogenous sums:
\begin{equation}\begin{split}\label{eq:544}
a_0 + b_1 &= \lambda (a_0 + b_0) + (1-\lambda) (a_1 + b_1) \\
a_1 + b_0 &= (1-\lambda) (a_0 + b_0) + \lambda  (a_1 + b_1)
,
\end{split}\end{equation}
with
\begin{equation}\label{eq:550}
\lambda = \frac{a_1 - a_0}{a_1 - a_0 + b_1 - b_0} \in [0,1]
.
\end{equation}
From \eqref{eq:544} and the convexity of $f$:
\begin{equation}\begin{split}\label{eq:544a}
f(a_0 + b_1) &\leq \lambda f(a_0 + b_0) + (1-\lambda) f(a_1 + b_1) \\
f(a_1 + b_0) &\leq (1-\lambda) f(a_0 + b_0) + \lambda  f(a_1 + b_1)
.
\end{split}\end{equation}
Summing the two equations in \eqref{eq:544a} yields the desired result \eqref{eq:536}.

The conclusion is that the sequence $N_{t-1}(x_{t})$ is non decreasing. Next, is it obvious that to increase the sum, all states in $\vr x$ should be chosen approximately the same number of times. If at some point, a state $x$ has been chosen for the $m$-th time at time $t$, while another state $x'$ did not appear $m-2$ times at this point, then because of the monotonicity of $N_{t-1}(x_{t})$, $x'$ can never appear again in an optimal sequence. Clearly, choosing $x'$ instead of $x$ at time $t$ would decrease $N_{t-1}(x_{t})$ for time $t$ as well as for all future occurrences of the state $x$, and therefore will increase the sum.

This concludes the proof that the last sum in \eqref{eq:342c} is maximized by $N_{t-1}(x_t) = \lfloor \frac{t-1}{|\mathcal{X}|} \rfloor$. Now, \eqref{eq:342c} may be rewritten as:
\begin{equation}\begin{split}\label{eq:342d}
\E [\hat L_n] - L^*_n
&\leq
\sum_{t=1}^n \frac{|\mathcal{X}| h_t - 1}{t}
+
\sum_{t=1}^n \frac{1}{\lfloor \frac{t-1}{|\mathcal{X}|} \rfloor + h_t}
\defeq \Delta
.
\end{split}\end{equation}
The last bound is only a function of $\{h_t\}$ and not of the sequence $\vr x$. If $h_t$ grows sublinearly, then the dominant factor in the second sum will be $\lfloor \frac{t-1}{|\mathcal{X}|} \rfloor$ and the sum would grow like $O(\log(n))$. On the other hand, any growth rate of $h_t$ that satisfies $\frac{1}{n} \sum_{t=1}^n \frac{h_t}{t} \arrowexpl{n \to \infty} 0$ would yield normalized expected regret tending to $0$, and a sufficient condition is $\frac{h_t}{t} \to 0$, and particularly this holds for $h_t = O(t^\alpha), \alpha \in [0,1)$.

If $h_t$ is constant, then the first sum grows like $O(\log(n))$ as well. A more detailed evaluation yields:
\begin{equation}\begin{split}\label{eq:342e}
\Delta
&=
(|\mathcal{X}| h - 1) \sum_{t=1}^n \frac{1}{t}
+
\sum_{t=1}^n \frac{1}{\lfloor \frac{t-1}{|\mathcal{X}|} \rfloor + h}
\\& \leq
(|\mathcal{X}| h - 1) \left( 1 + \int_{1}^n \frac{1}{t} dt \right)
+
\int_{t=0}^n \frac{1}{\frac{t}{|\mathcal{X}|} -1 + h} dt
\\&=
(|\mathcal{X}| h - 1) \left( 1 + \log(n)  \right)
+
|\mathcal{X}| \log \left(\frac{n - |\mathcal{X}|}{|\mathcal{X}| h} + 1 \right)
\\& \stackrel{h = |\mathcal{X}|^{-1}}{=}
|\mathcal{X}| \log \left(n - |\mathcal{X}| + 1 \right)
\\& \leq
|\mathcal{X}| \log (n)
.
\end{split}\end{equation}
This ends the proof of Theorem~\ref{theorem:FPF_log_loss}.

%Notice that this redundancy is similar to the redundancy obtained with Laplace's estimator and approximately twice the redundancy obtained using Kritchevsky-Trofimov's (which is approximately $\frac{|\mathcal{X}|-1}{2} \log n$). However notice that the target of the FPF forecaster was not to produce optimal redundancy for specific loss function.

%substituting
% d/dh (|\mathcal{X}| h - 1) \left( 1 + \log(n)  \right) + |\mathcal{X}| \log \left(\frac{n - |\mathcal{X}|}{|\mathcal{X}| h} + 1 \right)
% approx
% d/dh (|\mathcal{X}| h - 1) \left( 1 + \log(n)  \right) + |\mathcal{X}| (\log \left(\frac{n - |\mathcal{X}|}{|\mathcal{X}|} + 1 \right) - \log h)
% |\mathcal{X}| \left( 1 + \log(n)  \right) - |\mathcal{X}| / h = 0
% h ~= 1/log n

%The expression above allows us to compute an optimal value for the constant $h$ (TODO). Furthermore, optimizing $h_t$ per $t$ yields $h_t = \sqrt{\frac{t}{|\mathcal{X}|}} - \lfloor \frac{t-1}{|\mathcal{X}|} \rfloor$ which is not admissible (it is decreasing and sometimes negative).
%\frac{|\mathcal{X}|}{t} - \frac{1}{\left(\lfloor \frac{t-1}{|\mathcal{X}|} \rfloor + h_t\right)^2} = 0
%h_t = \sqrt{\frac{t}{|\mathcal{X}|}} - \lfloor \frac{t-1}{|\mathcal{X}|} \rfloor

\subsection{Proof of Theorem~\ref{theorem:Pu_is_universal}}\label{sec:proof_pu_universal}
Both Theorem~\ref{theorem:FPF_bounded_loss} and Theorem~\ref{theorem:FPF_log_loss} show the normalized expected regret tends to $0$ with $n$, and it remains to change from claims on expected regret to claims on the almost-sure regret. To prove that the regret tends to $0$ almost surely, or more precisely, $\limsup_{n \to \infty} (\hat L_n - L^*_n ) \leq 0$, using the already established fact that $\limsup_{n \to \infty} (\E [\hat L_n] - L^*_n ) \leq 0$, it remains to show that $\frac{1}{n} (\hat L_n - \E [\hat L_n]) \ntoinfty 0$ almost surely.
\begin{equation}\label{eq:415}
\frac{1}{n} (\hat L_n - \E [\hat L_n ])
=
\frac{1}{n} \sum_{t=1}^n \left( l(\hat b_t, x_t) - \E [ l(\hat b_t, x_t) ] \right)
\end{equation}
To show that the mean above converges to $0$ almost surely, we use Kolmogorov's criterion for the applicability of the Strong Law of Large numbers \cite{FellerProbability}. The elements of the sequence $\gamma_t = l(\hat b_t, x_t) - \E [ l(\hat b_t, x_t) ]$ have zero mean, and are independent, because each $\hat b_t$ depends only on the deterministic history of the sequence, and on $u_t(x)$, which are assumed independent. Notice that $\gamma_t$ are not identically distributed. Kolmogorov's criterion requires that $\sum_{t=1}^\infty \frac{\Var(\gamma_t)}{t^2} < \infty$. This holds trivially for bounded loss functions, for which the boundness of $|\gamma_t|$ yields a constant bound on its variance. Proving that this condition holds for the case of the log loss function is a rather technical calculation which is deferred to Appendix-~\ref{sec:proof_pu_universal_completion}.
\endofproof
%. For the log-loss, it was shown [***] that $\E [ l(\hat b_t, x_t) ]$ is bounded, and because $l(\hat b_t, x_t) \geq 0$, $\E \left| l(\hat b_t, x_t) - \E [ l(\hat b_t, x_t) \right| \leq  \E \left| l(\hat b_t, x_t) \right| + \left| \E [ l(\hat b_t, x_t) \right| = 2 \E l(\hat b_t, x_t) < \infty$. Using Theorem-XX in [REFREF] we have the desired result.
} % onlyfull (entire section)

\onlyfull{
\appendix
\subsection{Completion of the proof of Theorem~\ref{theorem:Pu_is_universal} for the log loss}\label{sec:proof_pu_universal_completion}
This appendix completes the proof of Theorem~\ref{theorem:Pu_is_universal} from Section~\ref{sec:proof_pu_universal}, by showing that for the case of the log loss, the Kolmogorov criterion holds and therefore the normalized regret converges almost surely to the normalized expected regret. Our purpose is to upper bound the following variance:
\begin{equation}\label{eq:807}
\sigma_t^2 = \Var(\gamma_t) = \Var(l(\hat b_t, x_t)) = Var \left[ \log \left( P^{(u)}_t(x_t) \right) \right]
,
\end{equation}
and show that Kolmogorov's criterion $\sum_{t=1}^\infty \frac{\sigma_t^2}{t^2} < \infty$ holds. $P^{(u)}_t(x_t)$ is defined in \eqref{eq:B109b} as
\begin{equation}\label{eq:811}
P^{(u)}_t(x) = \frac{N_{t-1}(x) + h_t \cdot u_t(x)}{t-1 + h_t \cdot \sum_{x' \in \mathcal{X}} u_t(x')}
.
\end{equation}
We have:
\begin{equation}\begin{split}\label{eq:817}
\log \left( P^{(u)}_t(x_t) \right)
&=
\log (N_{t-1}(x_t) + h_t \cdot u_t(x_t))
\\& \qquad - \log(t-1 + h_t \cdot \sum_{x' \in \mathcal{X}} u_t(x'))
\\&=
\underbrace{\log \left(\frac{N_{t-1}(x_t)}{h_t} + u_t(x_t) \right)}_A
\\& \qquad - \underbrace{\log \left(\frac{t-1}{h_t} + \sum_{x' \in \mathcal{X}} u_t(x') \right)}_{B}
.
\end{split}\end{equation}
Using
\begin{equation}\label{eq:821}
\Var(A-B) \leq E \left[ (A-B)^2 \right] \leq 2 E \left[ A^2 \right] + 2 E \left[ B^2 \right]
,
\end{equation}
where the second inequality stems from $(a-b)^2=2a^2+2b^2-(a+b)^2 \leq 2a^2+2b^2$, it is enough to bound the expected squared value of each $\log$ in \eqref{eq:817} separately. For a uniform r.v. $U \sim \unif[0,1]$ and a constant $a \geq 0$, the following bound holds:
\begin{equation}\begin{split}\label{eq:805}
\E \left[ \log^2(a + U) \right]
&=
\int_0^1 \log^2(a + u) du
\\&=
\int_{a}^{a+1} \log^2(u) du
\\&\leq
\int_{0}^{1} \log^2(u) du + \int_{\max(a,1)}^{a+1} \log^2(u) du
\\&\leq
\int_{0}^{1} \log^2(u) du + \int_{\max(a,1)}^{a+1} \log^2(u) du
\\&\leq
\left[ u \log^2(u) - 2 u \log(u) + u \right]_0^1 + \log^2(a+1)
\\&=
2 + \log^2(a+1)
.
\end{split}\end{equation}
We used the fact that $\log^2(u)$ is increasing for $u \geq 1$. Notice that the bound is trivial for $a \geq 1$ because $U \leq 1$. Applying the bound to the squared elements in \eqref{eq:817}:
\begin{equation}\begin{split}\label{eq:849}
\E \left[ \log^2 \left(\frac{N_{t-1}(x_t)}{h_t} + u_t(x_t) \right) \right]
& \leq
2 + \log^2 \left(\frac{N_{t-1}(x_t)}{h_t} + 1 \right)
\\& \leq
2 + \log^2 \left(\frac{t-1}{h_t} + 1 \right)
.
\end{split}\end{equation}
For the second term, the expectation is first applied only to one arbitrary element $u_t(x)$, while conditioning on the other elements:
\begin{equation}\begin{split}\label{eq:854}
&
\E \left[ \log^2 \left(\frac{t-1}{h_t} + \sum_{x' \in \mathcal{X}} u_t(x') \right) \right]
\\&=
\E \left\{ \E \left[ \log^2 \left(\frac{t-1}{h_t} + \sum_{x' \in \mathcal{X}} u_t(x') \right) | u_t(x'), x' \neq x \right] \right\}
\\& \stackrel{\eqref{eq:805}}{\leq}
\E \left\{ 2 + \log^2 \left(\frac{t-1}{h_t} + \sum_{x' \neq x} u_t(x') + 1 \right) \right\}
\\& \leq
2 + \log^2 \left(\frac{t-1}{h_t} + |\mathcal{X}| \right)
,
\end{split}\end{equation}
where we used again the fact that $\log^2(u)$ is increasing for $u \geq 1$. Combining \eqref{eq:817} with \eqref{eq:821} and the bounds above yields:
\begin{equation}\begin{split}\label{eq:870}
\sigma_t^2
&=
\Var \left[ \log \left( P^{(u)}_t(x_t) \right) \right]
\\& \leq
4 + 2 \log^2 \left(\frac{t-1}{h_t} + 1 \right) + 4 + 2 \log^2 \left(\frac{t-1}{h_t} + |\mathcal{X}| \right)
\\& \leq
8 + 4 \log^2 \left(\frac{t-1}{h_t} + |\mathcal{X}| \right)
\end{split}\end{equation}
Under the assumptions of Theorem~\ref{theorem:Pu_is_universal}, $h_t = h_1 \cdot t^{\alpha}$ with $\alpha \in (0,1)$ and so $\sigma_t^2 = O(\log^2(t))$ and clearly Kolmogorov's criterion holds.
\endofproof
} %onlyfull


\end{document}